\documentstyle[floats,prl,aps]{revtex}
\begin{document}
\renewcommand{\thefootnote}{\fnsymbol{footnote}}
\draft
\title{\large\bf 
         Bethe ansatz equations for Bariev's correlated
		 electron chain with boundaries}

\author{Yao-Zhong Zhang \footnote {Queen Elizabeth II Fellow.
                                   E-mail: yzz@maths.uq.edu.au}
             and 
        Huan-Qiang Zhou \footnote {On leave of absence from Dept of
	         Physics, Chongqing University, Chongqing 630044, China.
                 E-mail: hqzhou@cqu.edu.cn}} 

\address{      Department of Mathematics,University of Queensland,
		     Brisbane, Qld 4072, Australia}

\maketitle

\vspace{10pt}

\begin{abstract}
The Bethe ansatz equations are presented for Bariev's correlated
electron  chain with boundaries. This is achieved by using the 
coordinate space Bethe ansatz method.
\end{abstract}




\def\a{\alpha}
\def\b{\beta}
\def\d{\delta}
\def\e{\epsilon}
\def\g{\gamma}
\def\k{\kappa}
\def\l{\lambda}
\def\o{\omega}
\def\t{\theta}
\def\s{\sigma}
\def\D{\Delta}
\def\L{\Lambda}


\def\beq{\begin{equation}}
\def\eeq{\end{equation}}
\def\bea{\begin{eqnarray}}
\def\eea{\end{eqnarray}}
\def\ba{\begin{array}}
\def\ea{\end{array}}
\def\no{\nonumber}
\def\le{\langle}
\def\re{\rangle}
\def\lt{\left}
\def\rt{\right}

This work concerns with  Bariev's correlated electron
model \cite{Bar91} with boundary terms. 
The quantum integrability of this boundary model has been established by one
of the present authors \cite{Zhou96} using Sklyanin's formalism for the
boundary quantum
inverse scattering method \cite{Skl88}. The aim of this report is to diagonalize
this model by means of the coordinate space Bethe ansatz technique,
which has been used to solve the Hubbard open chain
\cite{Sch85,Asa96,Shi97} and the supersymmetric $U$ model \cite{Bra95} 
with boundary terms \cite{Zha97}.
We will derive the Bethe ansatz equations. Our results may be used to
investigate the finite size corrections to the low-lying energies in
the system. 

Our starting point is the Hamiltonian of Bariev's correlated electron
model with boundaries:
\bea
H&=&H^{\rm bulk}+H^{\rm boundary},\no\\
H^{\rm bulk}&=&\sum _{j=1}^{L-1}[(c^{\dagger}_{j\uparrow} c_{j+1\uparrow} +
 c^{\dagger}_{j+1 \uparrow}c_{j\uparrow})\exp(\eta n_{j+1 \downarrow})
    +(c^{\dagger}_{j\downarrow} c_{j+1\downarrow} + 
    c^{\dagger}_{j+1 \downarrow}c_{j\downarrow})\exp(\eta n_{j \uparrow})]
    ,\no\\
H^{\rm boundary}&=&{c_+ \exp(3\eta)}[2\sinh \eta n_{N\uparrow}n_{N\downarrow} 
    +\exp (-\eta)
    ( n_{N\uparrow}+n_{N\downarrow})]\no\\
& &     +\frac {1}{c_- \exp \eta}[2\sinh \eta n_{1\uparrow}n_{1\downarrow}
    +\exp(-\eta)
    ( n_{1\uparrow}+n_{1\downarrow})].\label{h}
\eea
Here $c^{\dagger}_{j\alpha}$ and $c_{j\alpha} $ are, respectively, the creation 
and annihilation operators of
electrons with spin $\alpha(=\uparrow $ or $\downarrow) $ at site $j$ and 
$n_{j\alpha}$ is the number density operator.

In order to derive the Bethe ansatz equations for the model, we
assume that the eigenfunction of Hamiltonian (\ref{h}) takes the
form
\bea
| \Psi \rangle & =&\sum _{\{(x_j,\s _j)\}}\Psi _{\s_1,\cdots,\s_N}
  (x_1,\cdots,x_N)c^{\dagger}
  _{x_1\s_1}\cdots c^{\dagger}_{x_N\s_N} | 0 \rangle,\no\\
  \Psi_{\s1,\cdots,\s N}(x_1,\cdots,x_N)
&=&   \sum _P \e _P A_{\s_{Q1},\cdots,\s_{QN}}(k_{PQ1},\cdots,k_{PQN})
  \exp (i\sum ^N_{j=1} k_{P_j}x_j),
\eea
where the summation is taken over all permutations and negations of
$k_1,\cdots,k_N,$ and $Q$ is the permutation of the $N$ particles such that
$1\leq   x_{Q1}\leq   \cdots  \leq  x_{QN}\leq   L$.
The symbol $\e_P$ is a sign factor $\pm1$ and changes its sign
under each 'mutation'. Substituting the wavefunction into the the
eigenvalue equation $ H| \Psi  \rangle = E | \Psi \rangle $,
one may get
\bea
A_{\cdots,\s_j,\s_i,\cdots}(\cdots,k_j,k_i,\cdots)&=&S_{ij}(k_i,k_j)
    A_{\cdots,\s_i,\s_j,\cdots}(\cdots,k_i,k_j,\cdots),\no\\
A_{\s_i,\cdots}(-k_j,\cdots)&=&s^L(k_j;p_{1\s_i})A_{\s_i,\cdots}
    (k_j,\cdots),\no\\
A_{\cdots,\s_i}(\cdots,-k_j)&=&s^R(k_j;p_{L\s_i})A_{\cdots,\s_i}(\cdots,k_j),
\eea
where $S_{ij}$ is the two-particle scattering matrix, and $s^L$ and $s^R$
are the boundary scattering matrices:
\bea
S_{ij}(k_i,k_j)&=& \left ( \begin {array} {cccc}
1 & 0 & 0 & 0\\
0 &\frac {\sinh (i \frac {k_1 -k_2}{2})}
{\sinh (i \frac {k_1 -k_2}{2}+\eta)}& 
\frac {e^{i\frac {k_1 -k_2}{2}}
\sinh \eta }{\sinh (i \frac {k_1 -k_2}{2}+\eta)} 
&0\\
0 &
\frac {e^{-i\frac {k_1 -k_2}{2}}
\sinh \eta }{\sinh (i \frac {k_1 -k_2}{2}+\eta)} 
&\frac {\sinh (i \frac {k_1 -k_2}{2})}
{\sinh (i \frac {k_1 -k_2}{2}+\eta)}& 
0\\
0 & 0 & 0 & 1
\end {array} \right ),\no\\ 
s^L(k_j;p_{1\s_i})&=&\frac  {1-p_{1\s_i}e^{ik_j}}
{1-p_{1\s_i}e^{-ik_j}}, \no\\
s^R(k_j;p_{L\s_i})&=&\frac {1-p_{L\s_i}e^{-ik_j}}
{1-p_{L\s_i}e^{ik_j}}e^{2ik_j(L+1)}.
\eea
In the above equations,
\beq
p_{1\s}\equiv p_1=c_-^{-1}e^{-2\eta},~~~~~~
p_{L\s}\equiv p_L=c_+ e^{2\eta}.
\eeq
For later use,let us introduce the notation
\beq
s(k;p)=\frac  {1-pe^{-ik}}
{1-pe^{ik}}.
\eeq
Then, the boundary scattering matrices $s^L$ and $s^R$ may be expressed
as
\bea
s^L(k_j;p_{1\s_i})&=&s(-k_j;p_{1\s_i}),\no\\
s^R(k_j;p_{L\s_i})&=&s(k_j;p_{L\s_i})e^{ik_j2(L+1)}.\label{sL-sR}
\eea
As is seen from the above equations, the two-particle $S$ matrix is 
nothing but the R-matrix of the asymmetric six-vertex model,
which satisfies the quantum Yang-Baxter equation (QYBE),
\beq
S_{ij}(k_i,k_j)S_{il}(k_i,k_l)S_{jl}(k_j,k_l)=
S_{jl}(k_j,k_l)S_{il}(k_i,k_l)S_{ij}(k_i,k_j)
\eeq
whereas the boundary scattering matrices $s^L$ and $s^R$ are
propotional to unit matrices, which satisfy the reflection equations,
\bea
&&S_{ji}(-k_j,-k_i)s^L(k_j;p_{1\s_j})S_{ij}(-k_i,k_j)s^L(k_i;p_{1\s
  _i})\no\\
&&~~~~~~~~~~~~~~~~~~=s^L(k_i;p_{1\s _i})S_{ji}(-k_j,k_i)s^L(k_j;p_{1\s _i})
  S_{ij}(k_i,k_j),\no\\
&&S_{ji}(-k_j,-k_i)s^R(k_j;p_{L\s_j})S_{ij}(k_i,-k_j)s^R(k_i;p_{L\s
  _i})\no\\
&&~~~~~~~~~~~~~~~~~~= s^R(k_i;p_{L\s _i})S_{ji}(k_j,-k_i)s^R(k_j;p_{L\s _i})
  ;p_{\s_i})S_{ji}(k_j,k_i).\label{reflection-e}
\eea
Indeed, substituting (\ref{sL-sR}) into (\ref{reflection-e}),
one obtains the consistency contition
for the matrix $s(k_j;p_{\s_i})$,
\bea
&&S_{ij}(k_i,k_j)s(k_i;p_{\s_i})S_{ji}(k_j,-k_i)s(k_j;p_{\s_j})\no\\
&&~~~~~~~~~~~~~~~~~~~~=
  s(k_j;p_{\s_j})S_{ij}(k_i,-k_j)s(k_i;p_{\s_i})S_{ji}(-k_j,-k_i).
\eea	
This is just the reflection equation discussed in
Mezincescu and Nepomechie's work \cite{Mez91} (see also
\cite{Zhou96}), which admits a trivial solution $s\propto 1$. 
Then, the diagonalization 
of Hamiltonian (\ref{h}) is reduced
to solving  the following matrix  eigenvalue equation
\beq
T_j\,t=t,~~~~~~~j=1,\cdots,N,
\eeq
where $t$ denotes an eigenvector on the space of the spin variables
and $T_j$ takes the form
\beq
T_j=S_j^-(k_j)s^L(-k_j;p_{1\s_j})R^-_j(k_j)R^+_j(k_j) 
    s^R(k_j;p_{L\s_j})S^+_j(k_j)
\eeq
with
\bea
S_j^+(k_j)&=&S_{j,N}(k_j,k_N) \cdots S_{j,j+1}(k_j,k_{j+1}),\no\\
S^-_j(k_j)&=&S_{j,j-1}(k_j,k_{j-1})\cdots S_{j,1}(k_j,k_1),\no\\
R^-_j(k_j)&=&S_{1,j}(k_1,-k_j)\cdots S_{j-1,j}(k_{j-1},-k_j),\no\\
R^+_j(k_j)&=&S_{j+1,j}(k_{j+1},-k_j)\cdots S_{N,j}(k_N,-k_j).
\eea
This problem may be solved using the algebraic Bethe Ansatz method.
Indeed, it can be shown that $[T_i,T_j]=0$ for any $i,\;j=1,\cdots,N$, 
which implies that 
$T_j,j =1,2,\cdots,N$ are simultaneously diagonalizable. We will not
present the proof here, but mention that 
$T_j$ is related to Sklyanin's transfer matrix $\tau(\l)$
for an inhomogeneous asymmetric six-vertex model with open boundary:
\bea
T_j&=&\frac{e^{-ik_j2(L+1)}}{\xi(k_j;p_1)\xi(k_j;p_L)}\;
      \tau (\l =k_j),\no\\
\tau (\l)& =&tr_0 (MT_{0}(\l)T_0^{-1}(-\l)),
\eea
where
\bea
\xi (k;p)&= &\frac {1-pe^{-ik}}{1-pe^{ik}},\no\\
T_0(\l)&=&S_{0N}(\l-k_N) \cdots S_{01}(\l-k_1),\no\\
M&=&\frac {\sinh (i\l +\eta)}{\sinh (i \l +2\eta)}\left ( \begin {array} {cc}
e^{-\eta} & 0\\
0 & e^{\eta}
\end {array} \right ).
\eea
Therefore, the commutativity of the
Sklyanin's transfer matrix $\tau(\l)$ for different values of $\l$
guarantees the commutativity of $T_j$'s. Diagonalizing $\tau(\l)$
using the algebraic (or analytical) Bethe ansatz method
and setting $\l=k_j$, we derive the following Bethe ansatz equations
\bea
e^{ik_j2(L+1)}\xi(k_j;p_1)\xi(k_j;p_L)
&=&\prod ^M_{\beta =1}\frac {\sinh (i\frac {k_j-v_\a}{2}+\frac {\eta}{2})
  \sinh (i \frac {k_j+v_\a}{2}+\frac {\eta}{2})}
 {\sinh (i\frac {k_j-v_\a}{2}-\frac {\eta}{2})
  \sinh (i \frac {k_j+v_\a}{2}-\frac {\eta}{2})}, \no\\
\prod ^N_{j =1}\frac {\sinh (i \frac {v_\a-k_j}{2}+\frac {\eta}{2})
  \sinh (i\frac {v_\a +k_j}{2}+\frac {\eta}{2})}
 {\sinh (i \frac {v_\a-k_j}{2}-\frac {\eta}{2})
  \sinh (i\frac {v_\a +k_j}{2}-\frac {\eta}{2})}
&=&
  \prod ^M_{\stackrel {\beta =1}{\beta \neq \alpha}}\frac
  {\sinh (i \frac {v_\a-v_\b}{2}+\eta)
  \sinh (i \frac {v_\a+v_\b}{2}+\eta)}
  {\sinh (i \frac {v_\a-v_\b}{2}-\eta)
  \sinh (i \frac {v_\a+v_\b}{2}-\eta)}.\label{bethe-eqn}
\eea
With the solutions $\{k_j\}$ of (\ref{bethe-eqn}), 
the energy eigenvalue $E$ of the model is given as
\beq
E=2\sum ^N_{j=1}\cos k_j .
\eeq

In summary, we have presented the Bethe ansatz equations for the 1D
Bariev's correlated electron chain with boundaries
in the context of the coordinate space Bethe ansatz technique.
Our results would be useful in analysing the structure of the ground
state and some low lying excitations of the model in the thermodynamic
limit. An interesting problem is to investigate spectrum of boundary
states in the model, as done in \cite{Bed97} for the open Hubbard
chain.

\vskip.1in
This work is supported by Australian Research Council, University of
Queensland New Staff Research Grant and External Support Enabling Grant. H.-Q.Z
would like to thank Department of Mathematics of UQ for kind hospitality.


\end{document}